# ELECTROMAGNETIC STRUCTURE FUNCTIONS OF NUCLEONS IN THE REGION OF VERY SMALL X

E.V. BUGAEV

*Institute for Nuclear Research of the Russian Academy of Sciences,
7a, 60th October Anniversary prospect, Moscow 117312, Russia*

B.V. MANGAZEEV†

*Irkutsk State University, 1, Karl Marx Street, Irkutsk 664003, Russia*

A two component model describing the electromagnetic nucleon structure functions in the low-x region, based on generalized vector dominance and color dipole approaches is briefly described.

## Introduction

A starting point of the present model of deep inelastic lepton-nucleon scattering is generalized vector dominance (GVD) idea [1], the main ingredient of which is a double dispersion relation for the total cross section for the process $\gamma^* p \to X$,

$$\sigma(s,Q^2) = \sum_q \int \frac{dM^2}{M^2+Q^2} \int \frac{dM'^2}{M'^2+Q^2} \rho(s,M^2,M'^2) \frac{1}{s} A_{q\bar{q}\to p}(s,M^2,M'^2). \tag{1}$$

Here, M and M' are invariant masses of the incoming and outgouing $q\bar{q}$-pair in $\gamma^* p$ elastic amplitude A.

It is assumed here that at high energies the virtual photon $\gamma^*$ fluctuates in a hadronic system ($q\bar{q}$-pair, to the leading order) well before the interaction with the target nucleon and, at the second step, $q\bar{q}$-pair interacts with the target. Later, in nineties [2-3], the relation (1) was used, in diagonal approximation, i.e., with $\rho \sim \delta(M^2 - M'^2)$, for a separation of the "soft" and "hard" interactions, and the soft part was described by the Vector Dominance Model (VDM).

The present two-component model of deep inelastic scattering differs (in a description of the soft part) from those developed in [2-3] in two respects: we do not use diagonal approximation and, most important, we do not limit ourselves by accounting of low mass vector meson (ρ) and, instead, take into account ρ-family (containing, except of ρ, the radial excitations ρ', ρ'',…). The use of the real mass spectrum of vector mesons is more natural than the introduction, in [3], the additive quark model component and, besides, allows introducing the non-diagonal transitions (and checking the importance of them).

## A two component model

In our previous papers [4], [5] we formulated the two-component model of electromagnetic structure functions of the nucleon. The nonperturbative (soft) component of the structure functions is described by the off-diagonal GVD with vector mesons having properties of usual hadrons. It was shown in [4] that the approach of the off-diagonal GVD alone cannot describe the experimental data if the destructive interference effects and corresponding cancellations of VN →V'N amplitudes inside of GVD sums are small (and they are really small if the vector mesons in the tower have the properties of usual hadrons). It was shown, as a result, that two modifications of the off-diagonal GVD scheme are needed: 1) cut-off factors reducing the probability of initial γ − V transitions must be introduced and 2) a "hard component" must be added to describe the perturbative QCD part of the total process of the virtual photon-nucleon interaction. For a calculation of the cut-off factor we use the following basic assumption: an interaction of the initial qq-pair with the nucleon is meson dominated if (and only if) the transverse momenta of pair's quarks are not too high; only in this case are confinement forces effective. It can be easily shown that the relative part of pair's phase volume for pairs with quarks having transverse momenta, $p^T$, in the limits ($m_q \div p^T_{max}$) is given, approximately, by $\eta \approx 3(p^T_{max}/M_{q\bar{q}})^2$, if $M^2_{q\bar{q}} >> (p^T_{max})^2$. By definition, this value is just the required cut-off factor. Here, the value $p^T_{max}$ is the model parameter. The average transverse size of the qq-pair is, at not very large $Q^2$, inversely proportional to $p^T$. From comparison with experimental data on $F_2$ we obtained for this parameter the value 0.385 GeV.

† Work partially supported by the program "Development of Scientific Potential in Higher Schools of Russia" (project 2.2.1.1/1483, 2.1.1/1539)".

The simplest model of VN-scattering had been used [4], based on two-gluon exchange approximation and relativistic constituent quark model. Wave functions of the vector mesons were obtained from a solution of the Bethe-Salpeter (BS) equation using quasipotential formalism in the light-front form. The kernel of the BS-equation has the confining term of the harmonic oscillator type. The vector meson mass spectrum is of the form $m_n^2 = m_\rho^2(1+2n)$, for the ρ-meson family (only this family has been taken into account). For the hard component, we used the color dipole model and the parameterizations of the dipole cross section $\sigma(r_\perp, s)$ (perturbative QCD part) from the work by J. Forshaw et al [6].

As one can see from fig.1, there is rather good agreement of the model predictions with the available data on the electromagnetic structure function $F_2$ in the region x< 0.05 and $Q^2 < 10^2$ GeV$^2$. The relative contribution in $F_2$ of the soft (GVD) component strongly depends on the values of the kinematic variables x, $Q^2$. At not very small x (x~$10^{-1} \div 10^{-2}$) the contribution of the soft component is dominant up to $Q^2$ ~ $10^2$ GeV$^2$. With a decrease of the x-value the interval of $Q^2$ in which soft component is dominant is reduced. For example, at x~ $10^{-4}$ the soft component is relatively large only in the region of very small $Q^2$, 0< $Q^2$ < 1GeV$^2$.

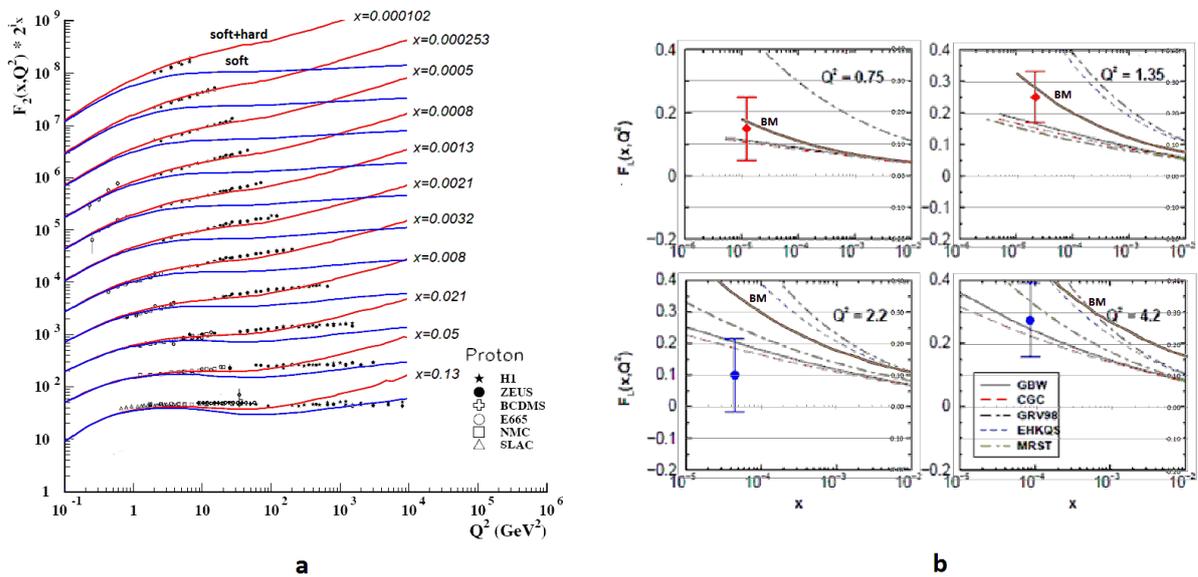

a

b

Figure 1. a) The proton electromagnetic structure function $F_2$ measured in experiments H1, ZEUS, BCDMS, E665, NMC, SLAC [7]. Data of each bin of fixed x has been multiplied by $2^i$, where i is the odd number of the bin, ranging from i=7 (x=0.13) to i=27 (x=0.000102). The red lines are predictions of the present model (soft+hard), blue lines are the contributions of the soft (GVD). b) The results for $F_L$ as a function of x at fixed $Q^2$. The data are taken from H1 Collaboration. Solid brown curves (BM) are results of the present model. Other curves are predictions of different theoretical models (see [8]).